%% file: main.tex
\documentclass{ieeeaccess}

\usepackage{color}
\usepackage{cite}
\usepackage{amsmath,amssymb,amsfonts}
\usepackage{algorithmic}
\usepackage{graphicx}
\graphicspath{ {images/} }
\usepackage{textcomp}
\def\BibTeX{{\rm B\kern-.05em{\sc i\kern-.025em b}\kern-.08em
    T\kern-.1667em\lower.7ex\hbox{E}\kern-.125emX}}

\usepackage{subcaption}
\usepackage{multirow}
\usepackage{float}
\usepackage{url} 
\usepackage{enumitem} 

\usepackage{textcomp}
\usepackage{gensymb}

\begin{document}

\history{Received April 26, 2019, accepted June 27, 2019, date of publication July 8, 2019, date of current version July 25, 2019.}
\doi{10.1109/ACCESS.2019.2927261}

\title{A parallel-computing algorithm for high-energy physics particle tracking and decoding using GPU architectures}

\author{
    \uppercase{Placido~Fernandez~Declara}\authorrefmark{1},\authorrefmark{2},
    \uppercase{Daniel~Hugo~C\'ampora~P\'erez}\authorrefmark{1},\authorrefmark{3},
    \uppercase{Javier~Garcia-Blas}\authorrefmark{2},
    \uppercase{Dorothea~Vom~Bruch}\authorrefmark{4},
    \uppercase{J.~Daniel~Garcia}\authorrefmark{2},
    and \uppercase{Niko~Neufeld}\authorrefmark{1},
}

\address[1]{EP-LBC, CERN, 1211--Geneve 23, Switzerland }
\address[2]{Department of Computer Science and Engineering, University Carlos III of Madrid, Madrid, Spain}
\address[3]{Universidad de Sevilla, ETSI Inform\'atica, Av. Reina Mercedes, s/n, 41012, Sevilla, Spain}
\address[4]{LPNHE, Sorbonne Universit\'e, Paris Diderot Sorbonne Paris Cit\'e, CNRS/IN2P3, Paris, France}

\corresp{Corresponding author: Placido Fernandez Declara (e-mail: placido.fernandez@cern.ch).}

\tfootnote{This work was supported by CERN; the European Research Council (ERC) under the European Union's Horizon 2020 research and innovation programme under grant agreement No 724777 "RECEPT"; and the Spanish MINISTERIO DE ECONOM\'IA Y COMPETITIVIDAD though project grant TIN2016-79637-P TOWARDS UNIFICATION OF HPC AND BIG DATA PARADIGMS.\hfill \copyright 2019 IEEE.  Personal use of this material is permitted.  Permission from IEEE must be obtained for all other uses, in any current or future media, including reprinting/republishing this material for advertising or promotional purposes, creating new collective works, for resale or redistribution to servers or lists, or reuse of any copyrighted component of this work in other works.}

\input{abstract_text}


\titlepgskip=-15pt

\maketitle

\input{sections/01introduction}
\input{sections/02related}
\input{sections/03background}
\input{sections/04ut_decoding_gpu}
\input{sections/05compass_ut}

\input{sections/06evaluation}
\input{sections/07conclusions}

\bibliography{bibliography.bib}{}
\bibliographystyle{IEEEtran}

\section{Acknowledgments}

The authors would like to thank Vladimir Gligorov and Florian Reiss for the fruitful discussions, Alberto Ottimo for early discussions on the UT decoding. We would also like to thank the LHCb computing and simulation teams for their support and for producing the simulated LHCb samples used to develop and benchmark our algorithm. Thanks to the ARCOS UC3M group for their support.

\EOD

\end{document}

%% file: abstract_text.tex
\begin{abstract}

Real-time data processing is one of the central processes of particle physics experiments which require large computing resources. The LHCb (Large Hadron Collider beauty) experiment will be upgraded to cope with a particle bunch collision rate of 30 million times per second, producing $10^9$ particles/s. 40 Tbits/s need to be processed in real-time to make filtering decisions to store data. This poses a computing challenge that requires exploration of modern hardware and software solutions. We present \emph{Compass}, a particle tracking algorithm and a parallel raw input decoding optimised for GPUs. It is designed for highly parallel architectures, data-oriented and optimised for fast and localised data access. Our algorithm is configurable, and we explore the trade-off in computing and physics performance of various configurations. A CPU implementation that delivers the same physics performance as our GPU implementation is presented. We discuss the achieved physics performance and validate it with Monte Carlo simulated data. We show a computing performance analysis comparing consumer and server grade GPUs, and a CPU. We show the feasibility of using a full GPU decoding and particle tracking algorithm for high-throughput particle trajectories reconstruction, where our algorithm improves the throughput up to 7.4$\times$ compared to the LHCb baseline.
\end{abstract}

%% file: sections/01introduction.tex
\section{Introduction}\label{sec:introduction}


High-energy physics experiments produce large data streams that must be processed, filtered, and analysed. The LHCb (Large Hadron Collider beauty) experiment is one of the four big physics detector experiments collecting data at the Large Hadron Collider (LHC). LHCb aims to explore the matter-antimatter asymmetry problem~\cite{Canetti_2012}. It is being upgraded and expected to restart operation in 2021; producing data at a rate of $40~Tbit/s$~\cite{framework2012upgrade}. Its event\footnote{A collision event corresponds to the crossing of two bunches of protons in the LHC beams.} filter will be run solely on general purpose computing resources, also known as software filter, where the LHCb data analysis framework has to process data in real-time, and decide which collision events may be discarded and which must be kept for further analysis. The software based event filter must be modernised to be able to handle the increased throughput~\cite{trigger2014online,lhcb2017upgrade}.

The LHCb experiment will have to increase its compute power needs to handle the continuous deluge of data from the detector. The big cost of the necessary increase in computer power lead to the exploration of alternative hardware architectures. As heterogeneous data centers comprised with multi- and many-core CPUs and coprocessors/accelerators emerge, LHCb and other CERN experiments are currently considering different hardware options to reach the aforementioned performance goals for the coming years. The current LHCb computing farm consists of servers based on the x86-64 architecture. However, alternative architectures and accelerators are being tested in different trigger systems~\cite{zhao2018new,vogelgesang2016high,vom2017online}. This is an indication that systems requiring high-throughput can be met in such alternative architectures.

LHCb computing farm needs to treat 30 million events per second, producing around $10^9$ particles per second. Reconstructing particle trajectories, known as particle tracking (from here on shortly referred to as "tracking"), plays a central role in processing these events. Introducing an architectural change, poses multiple challenges in terms of software to perform particle tracking in real-time. Existing algorithms must be redesigned to fully exploit parallel architectures. Furthermore, the expected long life cycle of these algorithms demands not only a high degree of performance optimization but also maintainability and portability. Those goals are ubiquitous in the scientific and engineering software areas and different solutions have been proposed. Among these, GPU-based approaches have been a successful alternative in providing high-throughput in different scenarios~\cite{NIETO2016957,Rogora2017subsetmatching,chen2018gpu}. This paper presents the implementation of a data-oriented approach, focusing on creating algorithms for SIMD (Single Instruction Multiple Data) architectures, minimizing thread divergence, reducing data movements and memory footprint of the algorithm, which have been successful strategies to optimize algorithms for GPUs~\cite{hsieh2016high,wang2016gunrock}. We run as part of the LHCb GPU sequence framework defined in~\cite{daniel2018unpublished}, which allows multiple concurrent GPU stream execution.

The main contributions of this paper are as follows:
\begin{enumerate}[label=\alph*)]
  \item We present a fast tracking algorithm for high-energy physics detectors targeting SIMD architectures called \emph{Compass}. The proposed algorithm can deal with deviated particle trajectories by a magnetic field.
  \item We introduce a parallel version for the decoding of the raw input data, which ensures coalesced data write patterns and produces a sorted SoA data structure,  beneficial to our tracking algorithm.
  \item We investigate the impact of our algorithm configuration on the physics quality of the results and analyze its computing performance on a variety of GPUs and CPUs.
\end{enumerate}

The rest of this paper is organised as follows. Section~\ref{sec:related} explores the state-of-the-art on high-throughput, real-time, and scientific usage of GPUs. Section~\ref{sec:background} briefly introduces the concepts used in high-energy physics for tracking, specifically for the LHCb experiment and UT tracking. On Section~\ref{sec:ut_decoding_gpu}, the implementation of the decoding of the raw input data is explained, whereas in Section~\ref{sec:compass_ut} the main algorithm design and implementation are presented. Section~\ref{sec:evaluation} shows the experimental evaluation carried out and presents the obtained physics efficiency. Finally Section~\ref{sec:conclusions} closes the paper with concluding remarks and future research lines.

%% file: sections/02related.tex
 \section{Related work}\label{sec:related}

We focus on high-throughput computing fields that process large scientific datasets and have similarities to those encountered in track reconstruction algorithms, this is, they process numerous small units of work. We discuss real-time approaches and other scientific applications which need to deliver high-throughput.

GPUs have been used before in the field of high-energy physics with success. The ALICE experiment at CERN implemented track reconstruction in GPUs obtaining different speedups compared to the previously used hardware~\cite{rohr2017gpu}. We note how the approach we follow is different than the one implemented in ALICE, as we aim to implement the full \emph{High Level Trigger} to run in GPUs, including the decoding and tracking of all subdetectors, thus avoiding much of the needed data transmission between main memory and GPU memory. Other HEP experiments have seen significant improvements when using GPUs to amend the performance of online selection~\cite{sen2015event, vom2017online}, or using a common code base to target both CPUs and GPUs using OpenCL, which shows the performance improvement of GPUs while supporting the x86-64 architecture~\cite{funke2014parallel}.

The performance of DNA sequencing problems has been improved with GPUs in different high-throughput scenarios. The \emph{Arioc} read aligner showed how using parallel algorithms with GPUs improved DNA sequencing throughput, achieving an order of magnitude faster alignments~\cite{wilton2015arioc, wilton2018arioc}. Pawar et al. benchmarked various DNA sequencing algorithms with different GPU-based tools against a CPU one; concluding that GPUs will replace CPUs in DNA sequencing for its higher-throughput processing~\cite{pawar2018evaluating}. Other DNA-related fields exhibit similar speedups: Samsi et al.~\cite{samsi2017linear} demonstrated how a single GPU is able to compare millions of DNA samples in seconds, Cadenelli et al.~\cite{cadenelli2019considerations} compared offloading a genomics workload into FPGAs and GPUs from a CPU, resulting in the GPU outperforming both, although the GPU consuming more energy.

Other scientific fields benefit from high-throughput, real-time processing in GPUs. Radio telescopes need to filter data in their data acquisition systems; where software frameworks employing GPUs like \emph{Bifrost}~\cite{cranmer2017bifrost} have shown significant performance improvements. Other real-time radio telescope experiments studied the viability of using GPUs, where they encountered large computing speedups at a local level, but were limited by I/O when using multiple GPUs~\cite{magro2014real}. Others in the same field have successfully implemented GPU optimization schemes~\cite{hu2018gridding} achieving a $6\times$ speedup compared to the CPU scenario, or used a GPU-based software framework and aggressive optimizations to be able to process data rates close to 1Tbit/s, like the \emph{CHIME Pathfinder} radio telescope~\cite{recnik2015efficient}.

GPUs have also been studied in scenarios requiring real-time processing at fusion experiments~\cite{maceina2017fast} greatly reducing the wall-time compared to the CPU version. Real-time split-and-merge executions have been improved in multi-GPU scenarios by Han et al.~\cite{han2016gpu}, and X-ray computer tomography reconstruction in GPUs has shown how different optimizations can be implemented and combined to speedup GPU computations~\cite{blas2014surfing}.

\begin{figure}[hbt!]
  \centering
    \includegraphics[width=\linewidth]{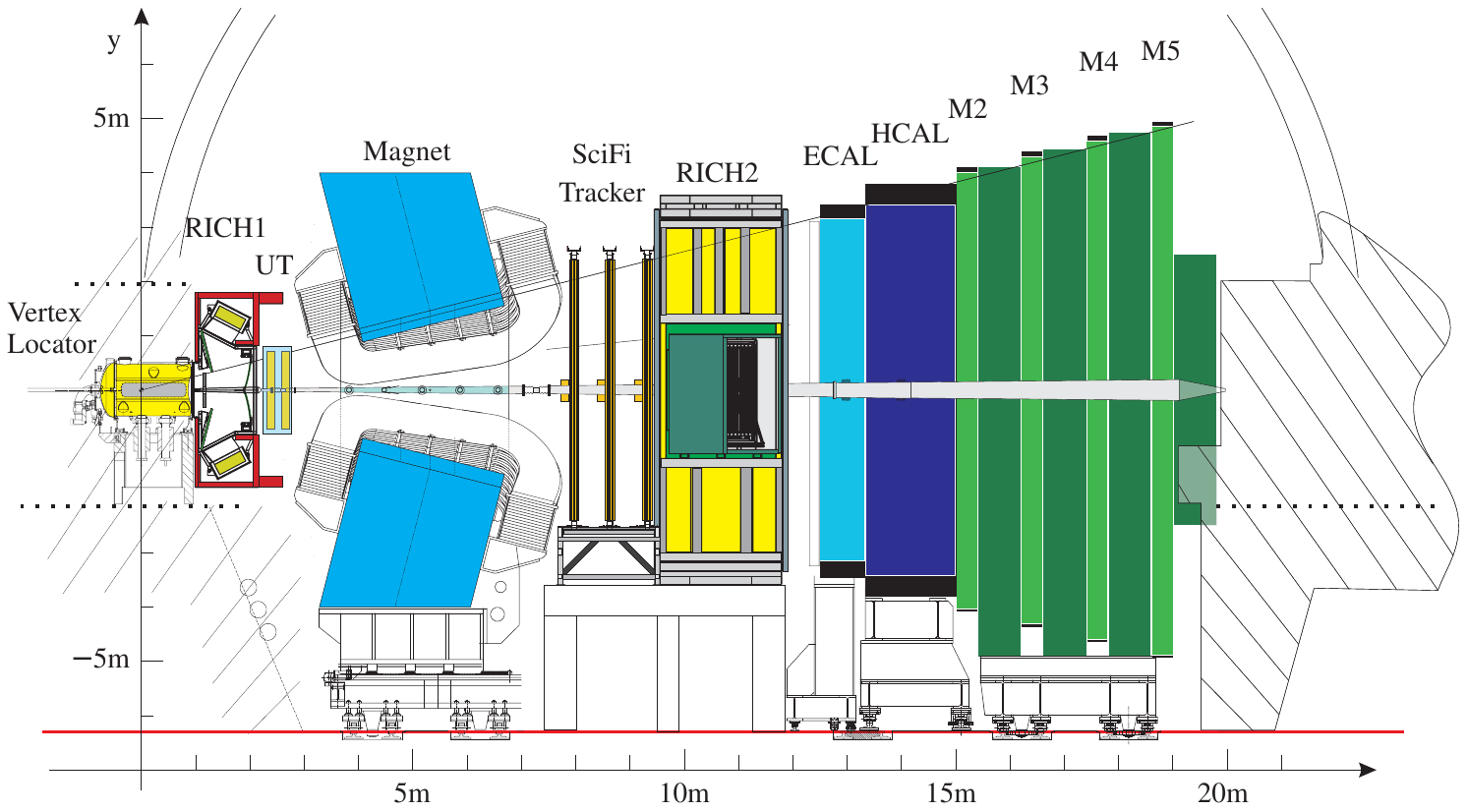}
  \caption{Schematic view of the LHCb upgrade detector.}
  \label{fig:lhcb_detector}
\end{figure}

Our approach for using GPUs in high-energy physics presents a parallel tracking algorithm which reconstruct particle trajectories that are bent under the influence of a magnet, describing a non-straight trajectory. We focus on achieving high-throughput to meet the collision rate and real-time constraints of the LHC at CERN. Other scientific fields have been successful on implementing real-time high-throughput solutions with GPUs, where fields like DNA sequencing are already ditching CPU-based architectures to process their large datasets. Successful results in the HEP fields suggest that implementing a full filter with GPUs, including the decoding and tracking of charged particles, is a feasible task that will increase the filtering throughput capabilities of LHCb.

%% file: sections/03background.tex
\section{Background}\label{sec:background}

\begin{figure*}[hbt!]
  \centering
    \includegraphics[width=0.9\linewidth]{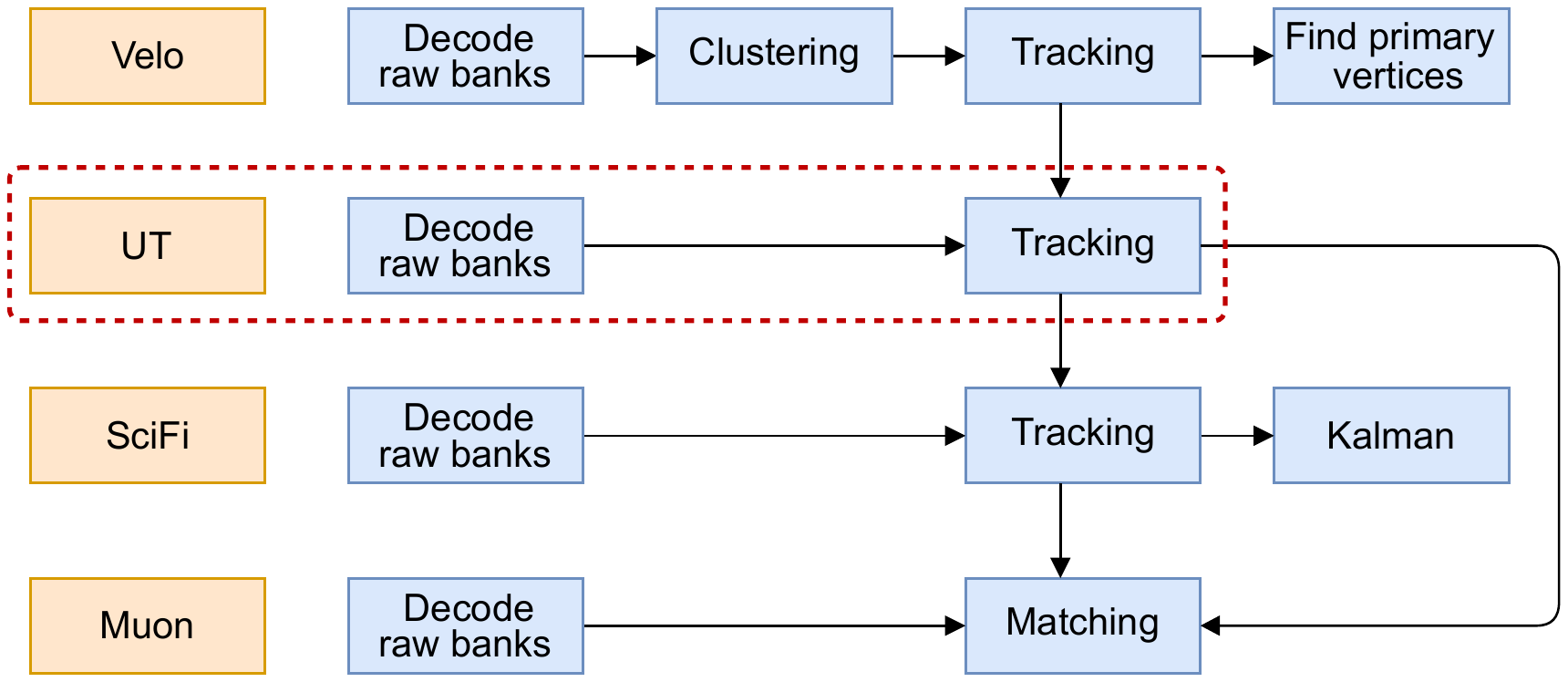}
  \caption{Complete High Level Trigger 1 sequence of algorithms at LHCb. We highlight the UT algorithms described in this paper (dotted lines). UT is the second tracking sub-detector in the chain of algorithms, and it receives input from the UT raw banks and the VELO tracks. UT outputs reconstructed tracks for other sub-detectors.}
  \label{fig:complete_chain}
\end{figure*}

In Figure~\ref{fig:complete_chain} we depict the full chain of algorithms needed to run the High Level Trigger 1 at LHCb required to filter events. In this section we describe the UT (Upstream Tracker) sub-detector, which provides part of the input data needed for the tracking algorithm. UT algorithms are second in the chain, receiving input from the UT raw banks and the reconstructed tracks from the VELO (Vertex Locator). This paper covers all the UT steps highlighted in Figure~\ref{fig:complete_chain}: the decoding of UT raw banks, and the UT tracking~\footnote{UT decoding and \emph{Compass} algorithms are available at \url{https://gitlab.cern.ch/lhcb-parallelization/Allen}}.

\subsection{UT sub-detector}\label{ssec:ut-sub-detector}

The LHCb detector is composed of various sub-detectors, as shown in Figure~\ref{fig:lhcb_detector}. In order to reconstruct particle trajectories, information from various sub-detectors is required. The sub-detectors that provide tracking information are the VELO, the UT, the SciFi Tracker and the $\mu$ (Muon) tracker. The UT is located in between the VELO and the SciFi Tracker~\cite{lhcb2017upgrade}. 

\begin{figure}[hbt!]
  \centering
    \includegraphics[width=0.9\linewidth]{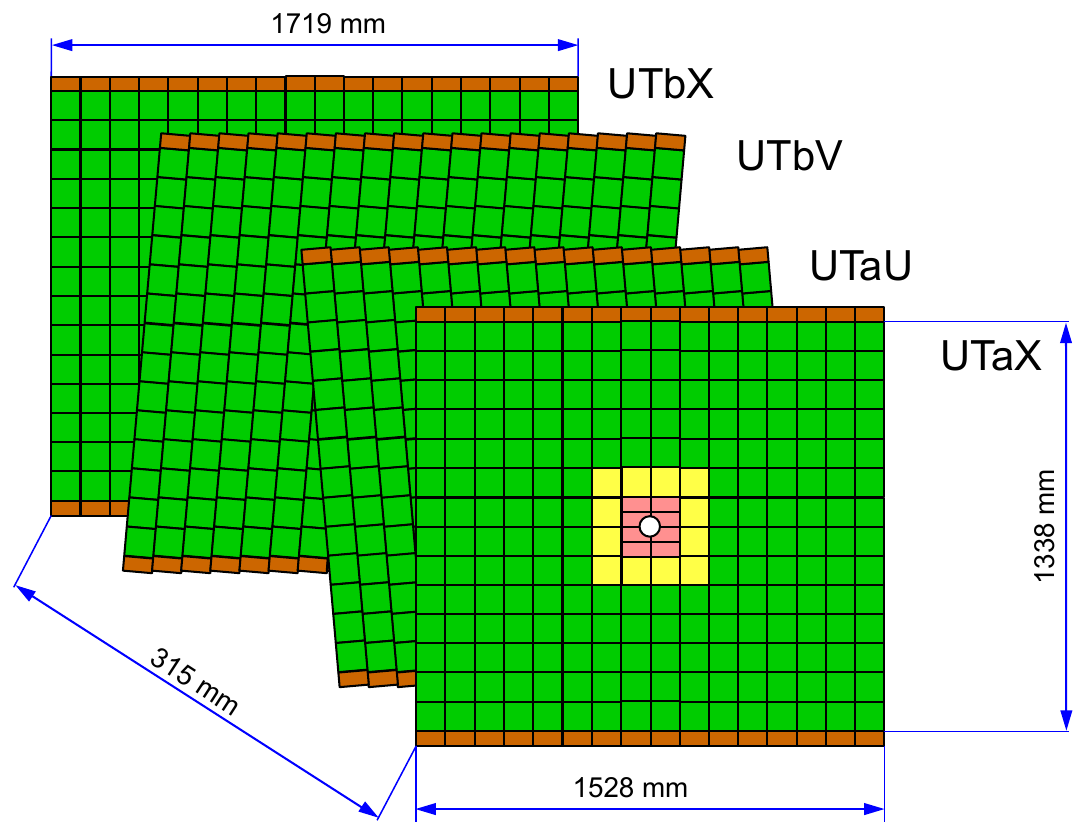}
  \caption{UT planes. The four UT planes are presented in this figure. The plane in the front (UTaX) is the closest to the VELO. Planes have a height of around 1.3~m, where the width is determined by the plane and changes between roughly 1.5 to 1.7~m. Different colors indicate the different types of sensors which accommodate different number of strips. The sensors around the centre have higher resolution. This design follows the simulation data, which indicates higher number of particles around the beam in the centre.}
  \label{fig:ut_planes}
\end{figure}

The UT sub-detector is composed of four planes, where each plane is a single sided silicon strip detector. We refer to the four consecutive planes as UTaX, UTaU, UTbV, UTbX respectively, as can be seen in Figure~\ref{fig:ut_planes}. These are sorted into two layers containing 2 planes each, the $a$ and $b$ layers. The $X$ planes are composed of vertical strips whereas the $U$ and $V$ planes are tilted around the Z axis at $+5\degree$ and $-5\degree$ respectively. By combining the measurements from the tilted $U$ and $V$ planes, the Y coordinate can also be determined. Each UT plane is composed of micro-strip sensors arranged in vertical staves~\cite{lhcb2014tracker}. A UT plane can be divided into 3 regions with different geometry, where the inner-most region has a finer granularity, and the outer regions have coarser granularity. Each stave measures 160~cm high and 10~cm wide, where various sensors are placed alongside each stave. The sensors in a stave overlap with their neighbour sensors, to avoid gaps, and the vertical staves also overlap for the same reason. The $X$ planes are composed of 16 staves while the $U$ and $V$ are composed of 18 staves. The acceptance of the UT sub-detector is defined by its volume in space, the UT planes for the UT sub-detector. Only particles that traverse this volume can leave signals and are measured.

The UT detector serves various purposes in the LHCb experiment:
\begin{itemize}[noitemsep]
    \item Reconstructs charged particles trajectories that decay after the VELO sub-detector.
    \item Reconstructs low momentum particles that are bent by the magnet, and go out of acceptance before reaching the SciFi Tracker.
    \item Gives additional information in the form of hits, that can be used in conjunction with the VELO and SciFi Tracker information to reject tracks.
    \item As the UT is influenced by the magnet, it can provide momentum resolution for charged particles.
    \item It can reject low momentum tracks.
    \item Decreases time to extrapolate VELO tracks to SciFi Tracker by at least a factor of 3.
\end{itemize}

Finally, UT plays an important role by marking tracks that won't be used by the next tracking detector, the SciFi Tracker. This allows for a faster processing of the whole track reconstruction in the LHCb detector.

\subsection{Track types, efficiency, and fake rates}

When performing particle tracking in the LHCb detector, tracks are classified according to the sub-detectors they traversed.

\begin{figure}[hbt!]
  \centering
    \includegraphics[width=0.9\linewidth]{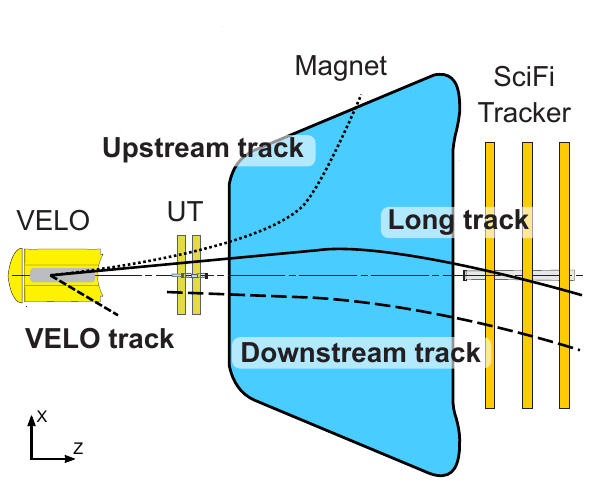}
  \caption{LHCb track types. Each track type is classified according to the sub-detectors it traverses. This figure represents a top view of the tracking sub-detectors, where particles travel from the collision point at the VELO to the right, crossing the UT and the SciFi Tracker, or travelling out of acceptance.}
  \label{fig:track_types}
\end{figure}

The tracks that traverse the UT sub-detector or serve as input for it are classified as follows:
\begin{itemize}
    \item \textit{Long tracks}: contains hits detected from the VELO to SciFi Trackers, and they may contain hits in the UT. Long tracks analysed here have hits in the UT.
    \item \textit{Upstream tracks}: comprise hits recorded in VELO and UT detectors, but not in SciFi Tracker. These tracks are bent by the magnetic field, so they travel outside the SciFi tracker, without crossing it. We refer to them as VELO+UT tracks.
    \item \textit{Downstream tracks}: contains hits recorded in UT and SciFi detectors, but not in the VELO, so their origin is external to the collision point. These tracks are not relevant for the tracking algorithm covered in this paper, but they leave hits in the UT sub-detector that are not matched to a VELO track.
    \item \textit{VELO tracks}: contains hits recorded in the VELO. During UT tracking, these tracks are extended to other types of tracks if matching hits are found.
\end{itemize}

In the context of the LHCb experiment, long tracks play an important role as they traverse the full magnetic field and therefore have the most precise momentum information~\cite{VanTilburg2005tracksimulation}. 

When doing the track reconstruction, a particle is considered to be \emph{reconstructible} in the UT sub-detector if it has hits in three of the four layers. Various parameters are measured to determine physics efficiency~\cite{schiller2011track}:
\begin{itemize}[noitemsep]
    \item \textit{Track reconstruction efficiency}: It is measured with simulation data comparing the number of tracks correctly reconstructed against the number of tracks that are reconstructible. To be considered \emph{reconstructed}, 70\% of the hits on a track need to be associated to the particle from the Monte Carlo simulation. The reconstruction efficiency is given as: 
    \\ \begin{center} $\dfrac{N_{reconstructed~\&~reconstructible}}{N_{reconstructible}}$ \end{center} 
    \vspace{1em}
    \item \textit{Clone rate}: When two or more tracks are associated to the same Monte Carlo particle, only one is considered to be reconstructed correctly and the others are counted as \emph{clones}. The clone rate is the number of clone tracks relative to all reconstructed tracks. The clone rate is defined as: 
    \\ \begin{center} $\dfrac{N_{clone~tracks}}{N_{reconstructed~tracks}}$ \end{center} \vspace{1em}
    \item \textit{Fake rate}: A track is considered a fake when it is reconstructed, but it cannot be associated with a Monte Carlo particle. The fake rate is defined as follows: 
    \\ \begin{center} $\dfrac{N_{fake~tracks}}{N_{reconstructed~tracks}}$ \end{center} \vspace{1em}
\end{itemize}

We refer to physics efficiency to describe how good our tracking algorithm is performing, analogous to a cost function that uses the three parameters, reconstruction efficiency rate, clone and fake rates. There is no analytical form of such cost function, where an algorithm is said to attain good physics efficiency if the reconstruction efficiency is high, and the clone and fake rates are low.

\subsection{UT tracking}\label{ssec:ut_tracking}

Particles collide at the interaction point, and the resulting particles from the collisions are first reconstructed by the VELO sub-detector. A percentage of those particles travel out of the acceptance range of the UT, and the rest of them, in acceptance, leave activation signals with a high probability which are decoded in software to hit information. Using the VELO tracks and the UT hit information, combined with the geometry information and magnetic field influence from the magnet, we are able to perform the UT tracking. 

\begin{figure}[hbt!]
  \centering
    \includegraphics[width=0.9\linewidth]{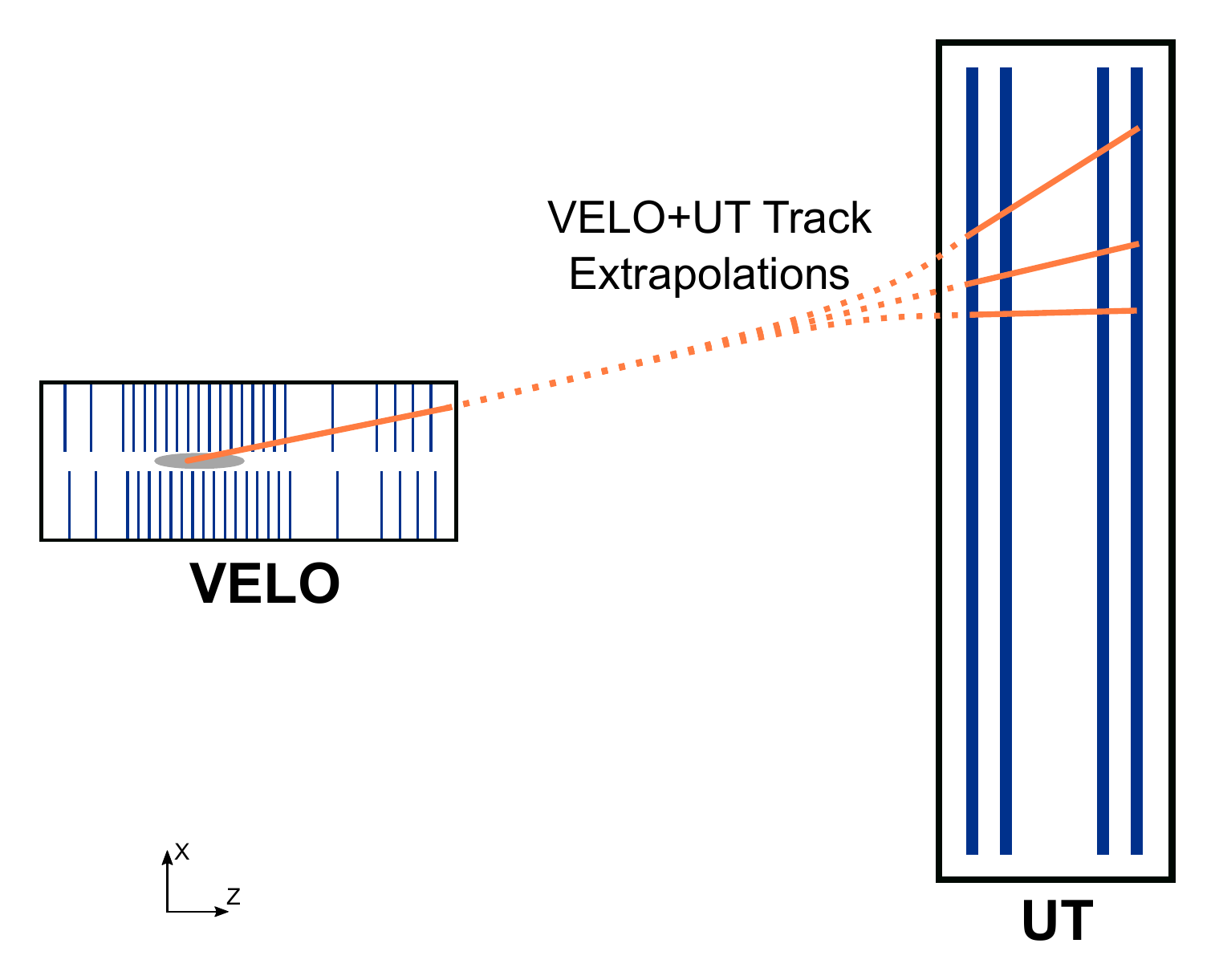}
  \caption{VELO track extrapolation to UT hits. A VELO track can be associated to various UT hits, where the UT track extrapolation does not necessarily follow a straight line. This leads to high combinatorics between the hits in the four panels, holding the main complexity of the algorithm.}
  \label{fig:ut_combinatorics}
\end{figure}

Tracking is done by finding matching UT hits for every input VELO track, where a VELO track is a straight line. UT hits are considered to be compatible with a VELO track, resulting in a curved track bent proportionally to the track momentum. As the UT sub-detector is under the influence of the magnetic field, multiple possible matching hits can be matched for different slightly bent tracks~\cite{chouaki2016upstreamtracking}. This situation is represented in Figure~\ref{fig:ut_combinatorics}, where a real situation is better represented with hundreds of tracks, and makes the problem of finding matching hits an exponential combinatorics problem~\cite{Bowen2016upstream}.

The p-Kick method~\cite{Bowen2014velout} is used to estimate the momentum of the track. Using it allows to perform a $\chi^2$ fit providing the momentum of the particle. This method is used instead of a Kalman filter, used in other tracking algorithms, as it yields a better computing performance~\cite{Bowen:2105078}. To take into account the magnetic field during the algorithm, look-up tables are used, which give quick access to the influence of the magnetic field in different parts of the particle trajectory. Using the look-up tables, the deflection a track is expected to experience can be determined.

A UT tracking algorithm is expected to achieve a high reconstruction efficiency with a low fake and clone rates for various types of tracks. The computing performance of the algorithm is determined by how many events per second can be processed for a given hardware configuration. This is a key aspect of event filtering in high-energy physics, especially for the LHCb experiment which will rely only on software for its event filter system. The combination of hardware and optimised software for it will need to process the 30~MHz rate of events in real-time.

%% file: sections/04ut_decoding_gpu.tex
\section{UT decoding on GPU}\label{sec:ut_decoding_gpu}

Before being able to execute the tracking algorithm, the raw input from the subdetector needs to be decoded into hit information. The decoding step needs to perform efficiently to run in real-time. We parallelise it by processing different chunks of raw input using GPUs, as it is a fundamental previous step for the tracking algorithm.

UT detector data is encoded into raw banks, in a highly compact format, containing the information required to obtain the UT hits. These raw banks are decoded into the parameters that define a UT~hit. We reduced the decoded parameters to the minimum to run the UT tracking algorithm, lowering the memory footprint of the algorithm. The decoded parameters are the following:
\begin{itemize}
    \item \textit{LHCbID}: a unique 32 bit identifier for the hit, which indicates the spatial position of the detection element.
    \item \textit{$Z$ at $Y=0$}: the $Z$ coordinate of the hit at the $Y=0$ position, which is the centre of the panel in the $Y$ axis. The $Z$ coordinate indicates the panel for a specific hit.
    \item \textit{$X$ at $Y=0$}: similarly to the previous parameter, this is the $X$ position at the centre of the panel in the $Y$ axis. This coordinate is given by the activated strip in a sector and it is different for the $U$ and $V$ layers.
    \item \textit{yBegin} and \textit{yEnd}: as the UT subdetector is a strip detector where the strips are arranged vertically, the specific $Y$ coordinate of a hit cannot be gotten. Instead, a range on the $Y$ axis delimits where the hit is located.
    \item \textit{weight}: the uncertainty of the hit position.
\end{itemize}

The decoded parameters are stored in a structure of arrays (SoA). We use a SoA layout storing the hits in a coalesced manner to maximise the memory bandwidth usage. To access the hits efficiently, a separated array is used to store the offsets between the hits. Using the offsets, we are able to determine which panel we are referring to when accessing the hits, and so every GPU thread can access its specific hit. We compute the events by processing them in parallel, assigning single events to single blocks to distribute them in the GPU. An event results in various tracks, where we apply different nested parallelisation schemes for different kernels, which are described here.

\begin{table}[hbt!]
\begin{center}
\caption{Kernel configuration for UT decoding. $events\_in\_execution$ are the number of selected events to process, where $array\_size$ is defined as the $events\_in\_execution \times 84$. 84 is the number of pre-defined sectors, where the number 4 used in various kernels is the number of panels.}
\label{table:kernel_decoding}
\resizebox{0.98\linewidth}{!}{
\begin{tabular}{l|l|c}
kernel                    & blocks                              & threads      \\
\hline
calculate number of hits  & $events\_in\_execution$             & (64,4)       \\
prefix sum reduce         & $(array\_size + 511) / 512$        & 256          \\
prefix sum single block   & 1                                   & 1024         \\
prefix sum single scan    & $((array\_size + 511) / 512 ) - 1$ & 512          \\
pre-decode                & $events\_in\_execution$             & (64,4)       \\
find permutation          & $(events\_in\_execution, 84)$       & 16           \\
decode raw banks in order & $(events\_in\_execution, 4)$        & 64           
\end{tabular}
}
\end{center}
\end{table}

We group the decoded hits into \emph{sector groups}, which are composed of various sensors. Each \emph{sector group} carries a number of hits that are guaranteed to be within certain $X$ coordinates. Within a \emph{sector group} hits are not sorted by $X$ coordinate, making it faster to sort. This also allows for quick look-up of hits in the tracking algorithm, targeting specific sector groups and searching hits only in those. Hits are sorted into pre-defined regions of the \emph{sector groups}, then sorted by $Y$ coordinate within the \emph{sector group}. We divide the complete decoding into 7 GPU kernels, where we found the configuration in Table~\ref{table:kernel_decoding} to be the fastest for the UT decoding.

\begin{itemize}
    \item \textit{Calculate number of hits}: the first kernel uses pre-defined regions in the $X$ axis, where the regions in the center of the panel are narrower due to the increased number of tracks expected based on previous LHCb data takings. Raw banks are processed to calculate the number of hits, used to create the array to store the offsets between the hits in memory, in a coalesced manner. To process the raw banks in parallel we set a two-dimensional kernel, parallelising over the raw banks and over the number of hits in each raw bank.
    \item \textit{Prefix sum}: we implement a parallel prefix sum of the hits, specifically a \emph{two-step Blelloch scan} composed of a reduce and down sweep operations. It results in an array with the sums of the offsets, so their positions and sizes can be obtained~\cite{harris2007parallel}. After doing the prefix sum the total number of hits is obtained, which allows us to pre-allocate the memory for the hits. The prefix sum is implemented here in three separate kernels, as seen in Table~\ref{table:kernel_decoding}.
    \item \textit{Pre-decode}: using the data structure created during the prefix sum, the coordinates of the hits for each raw bank can be decoded. Parallelising over the raw banks and over the number of hits in each raw bank, the strip information to get the subdetector region, panel and sector of the hit is extracted. Using this information we decode the \emph{X at Y=0}, and \emph{yBegin} coordinates to delimit the hit in the $Y$ axis.
    \item \textit{Find permutation}: it calculates the required permutations to sort the hits by $Y$ coordinate, based on their decoded $Y$ coordinate limits. Hits are sorted within every group defined by the previously decoded $X$ coordinate. We implement an insertion sort in shared memory, storing the $Y$ coordinate in it, and parallelising over the hits found in each sector group.
    \item \textit{Decode raw banks in order}: to perform the actual decoding of the UT hits a gather operation is used. It gets geometry and panel information from the subdetector, and stores the parameters in a coalesced manner. The hit information is stored in its correct position using the pre-defined $X$ coordinate regions and the permutations calculated in the previous kernel. For this kernel, we parallelise over the hits found on each layer.
\end{itemize}

%% file: sections/05compass_ut.tex
\section{Compass tracking algorithm}\label{sec:compass_ut}

We designed the \emph{Compass} tracking algorithm so it can be configured by two parameters: the number of sectors to search for hit candidates, and the number of valid found candidates to test to form a track. Different configurations of these parameters gives us a configurable trade-off between computing and physics performance.

\emph{Compass} is focused on the SIMD many-core parallelism offered by GPUs and its memory characteristics to develop a high-throughput algorithm. To achieve high-throughput we perform tracking on thousands of tracks in parallel, in real-time, where each particle trajectory can be computed independently one from each other. We benefit from this to design the algorithm around an SIMD model, where GPUs implement it in a SIMT (Single Instruction Multiple Thread) execution model. The operations needed to calculate the particle trajectories require arithmetic and matrix operations with single precision floating point numbers, where GPUs have shown to offer speed-ups in scientific computations. We access the decoded window ranges stored in a SoA data layout. Other multi-threaded architectures like modern x86-64 should also benefit from a SoA layout, as the access pattern by the different threads also benefit from data locality and coalesced access. The NVIDIA Profiler was used to optimize and find the spots to parallelize.

\emph{Compass} is divided in two main components: searching for the UT window ranges in the indicated sectors, and using those window ranges to perform the tracking. In both cases, VELO tracks are used as input, and are extrapolated to the UT panels.

\subsection{Search UT windows}\label{ssec:search_ut_windows}

UT window ranges are defined by the indexes of two hits, one at the beginning of the window and the other at the end, where hits in between these two are considered for creating a track. The search for UT windows is performed using the information about how hits are sorted during the decoding. A two-dimensional kernel is used to search the windows: the first dimension parallelises over the four UT panels, where the second does it over the input VELO tracks. We define the kernel like this to optimize for the windows ranges to be stored in SoA layout, where we tested different kernel configurations, concluding this one to yield the best performance. Window ranges are stored in a coalesced manner for a panel, where panels are also stored contiguously between them. The two-dimensional kernel is used to favour the access pattern, first over the panels, then over the different tracks. We found this configuration to be faster than setting the kernel the opposite way, or just parallelising over the tracks in a one-dimensional kernel.

\begin{figure}[hbt!]
  \centering
    \includegraphics[width=\linewidth]{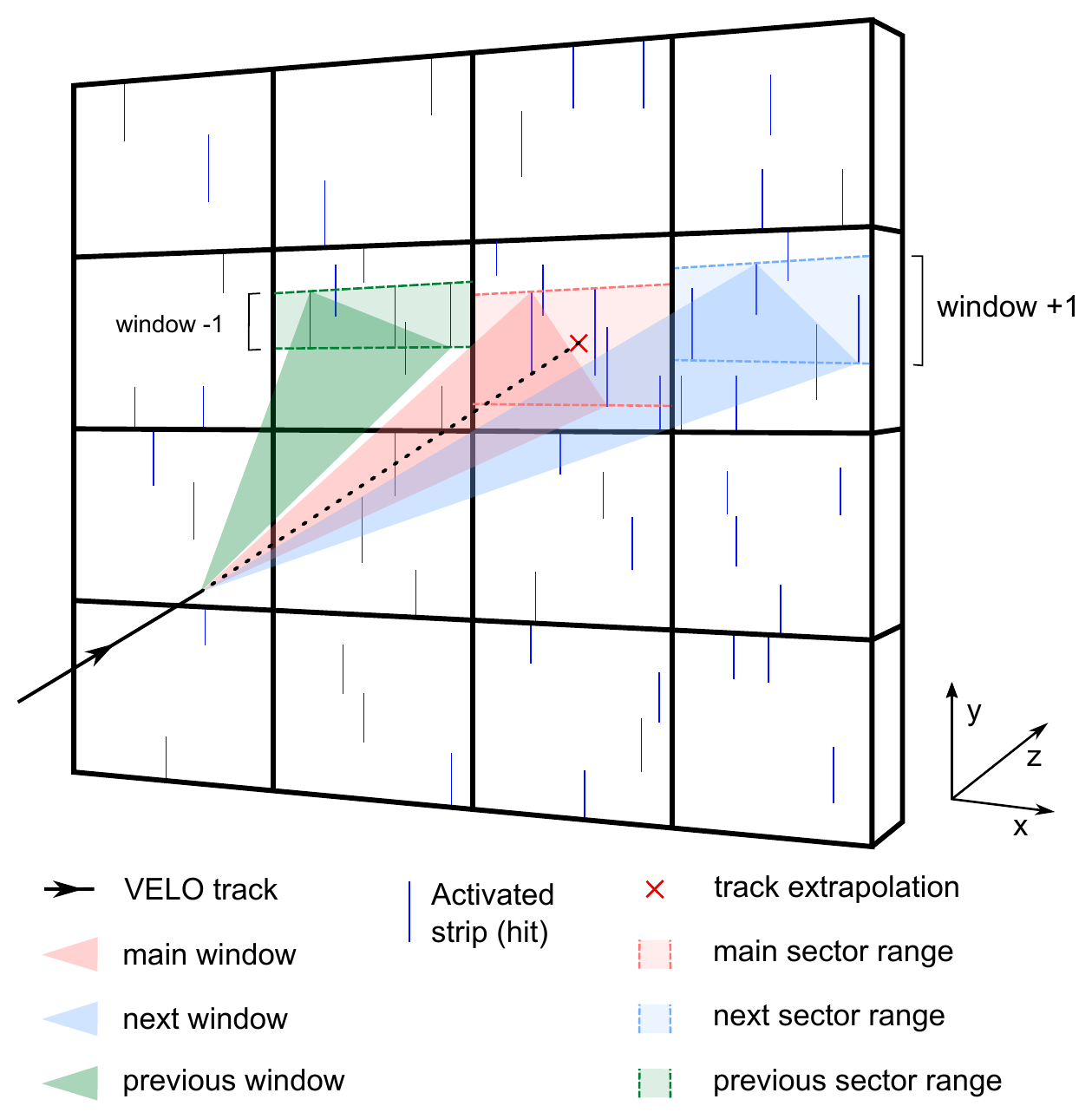}
  \caption{UT window ranges: representation of a VELO track extrapolation to a sector. Window ranges are set for the sector and its neighbours. Several hits lie within the range of the windows, which are considered for UT tracking}
  \label{fig:ut_windows}
\end{figure}

For each input VELO track, the extrapolation to the UT panels is calculated taking into account the magnetic field. The extrapolation defines the sector group in the UT to search for. Since sector groups are sorted by $X$ into known regions, a binary search is used to efficiently locate the region where the extrapolation is pointing to. With the region delimited by $X$, a tolerance window based on the VELO track extrapolation is used to delimit the $Y$ region. Searching with two binary searches over the $Y$ axis, one to delimit the beginning of the region and another to delimit the end of it, leaves us with the window range that indicates the valid UT hits for the associated VELO track. Only two pointers to the hits are used to indicate a window range. Finally the window range is refined by checking the hits to be valid within the VELO tolerance window. Iterating forward for the beginning hit, and backward for the end hit, hits are tested to meet the conditions for the VELO track tolerance. This calculation is performed here to reduce the window ranges, which we found to be faster compared to only perform it in the tracklet finding kernel. When computing the tracking kernel combinations between the hits in different panels are tested. Using a larger window range during the tracking has a larger impact in the complexity to compute the kernel compared to refining the window range during the window search. As the hits in a sector group are not sorted, the VELO tolerance check has to still be performed again in the tracking kernel because hits could be out of the tolerance window.

When looking for window ranges, a VELO track may be outside the UT acceptance region or may be directed in backwards direction, making the track unsuitable for UT tracking. When a thread is assigned to a track that meets any of those conditions, the whole thread is left unused until the rest of the threads in its \emph{warp} finish finding the window regions. Some threads are left unused for every event, lowering the throughput capacity of the algorithm. To maximize thread occupation an array of pointers to tracks in shared memory is used, which is filled with valid tracks only. The array is filled until it holds at least the same amount of tracks as number of threads per block. We search windows parallelising over the array of pointers to valid tracks, maximizing thread occupation.

\begin{figure*}[hbt!]
  \centering
    \includegraphics[width=\linewidth]{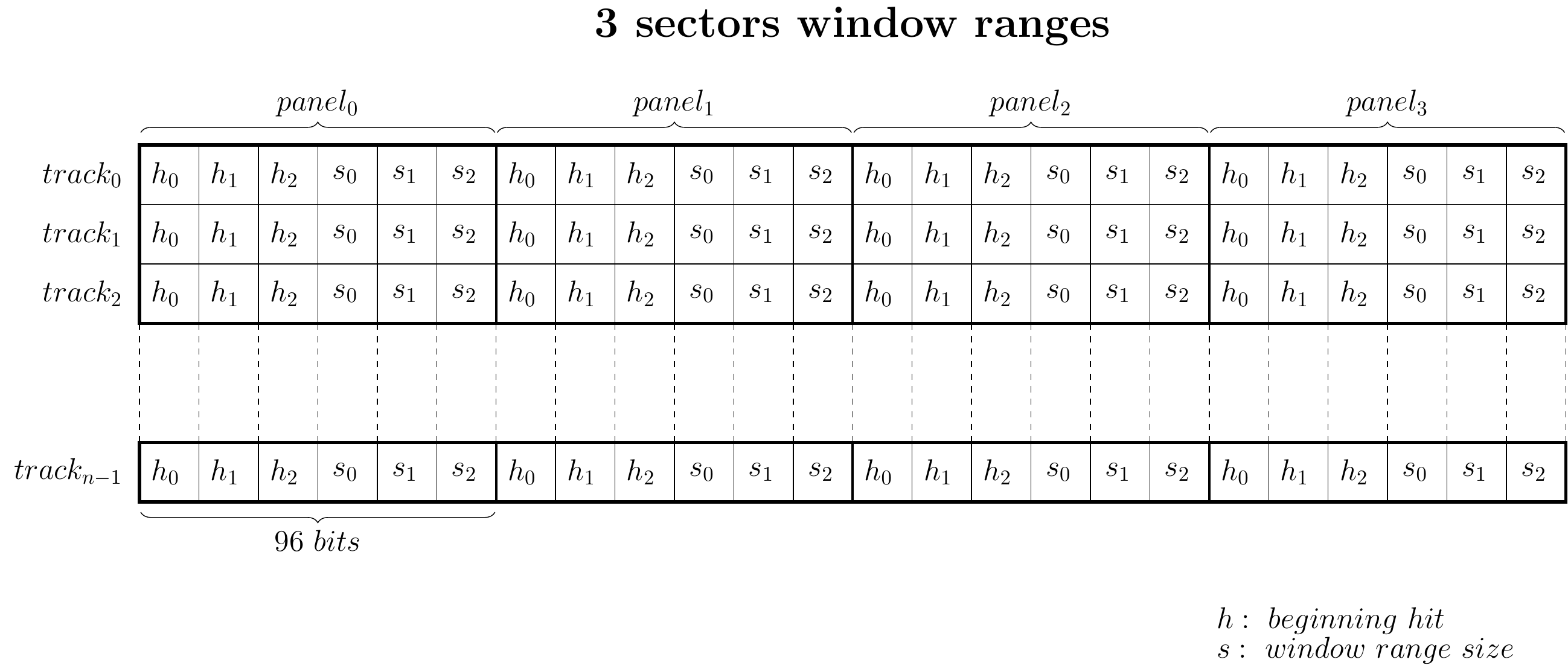}
  \caption{Memory layout of window ranges. A beginning hit, and a size are stored per window range, using 16 bits for each element. In this figure, a 3 sectors window ranges is shown, where each elements has a size of 16 bits, making it a total of 96 bits for all the elements of a panel.}
  \label{fig:memory_windows}
\end{figure*}


We implement the window search to look for hits in one, three or five sectors. We do this because we found the number of hits found in only one sector to be insufficient to achieve good enough physics performance. The selected sector and its neighbours are used to get hit candidates, as can be seen in the Figure~\ref{fig:ut_windows}. If the extrapolated VELO track is pointing to a sector close to the borders of the UT panel, less sectors are searched. The window ranges are stored in a pre-allocated memory space, as the number of sectors to use and VELO track is already known, so they can be stored in parallel for every thread. When an invalid window range is found, it is stored with \texttt{(-1, -1)}, indicating that no valid hits were found. By doing this the kernel presents a lower branching ratio, leaving a similar code path for all tracks searching the windows, making it efficient for GPUs.

Finally, window ranges are stored as pairs composed of a beginning hit and the size of the window. As we will iterate over the hits in the window, knowing in which window the hit starts and the size of the window is all the information we need to access the hits. To store the hit and the size of each window we use two signed 16-bit types (\texttt{short}). The hit index is set to be relative to its own track, for all the possible indexes to fit in a \texttt{short} type, thus reducing the memory footprint. Hit pointers and window range sizes are stored grouped so all hits are contiguous between them, and per track, as can be seen in Figure~\ref{fig:memory_windows}.

\subsection{Tracklet finding}\label{ssec:tracklet_search}

To perform UT tracking, a search for the best compatible hits needs to be performed in all the UT panels to form a tracklet. A tracklet is composed of at least 3 hits on different panels. The combination that best matches the extrapolation from the VELO track is searched, considering the influence of the magnetic field that introduces a small kink in the particle trajectory. The window ranges calculated in the previous kernel are used to find a tracklet of one hit per UT panel, allowing for one missing UT hit. The main complexity of \emph{Compass} lies in the tracklet search, where compatible hits between all panels are tested for compatibility, increasing the multiplicity of the combinations.

\begin{figure}[hbt!]
  \centering
    \includegraphics[width=\linewidth]{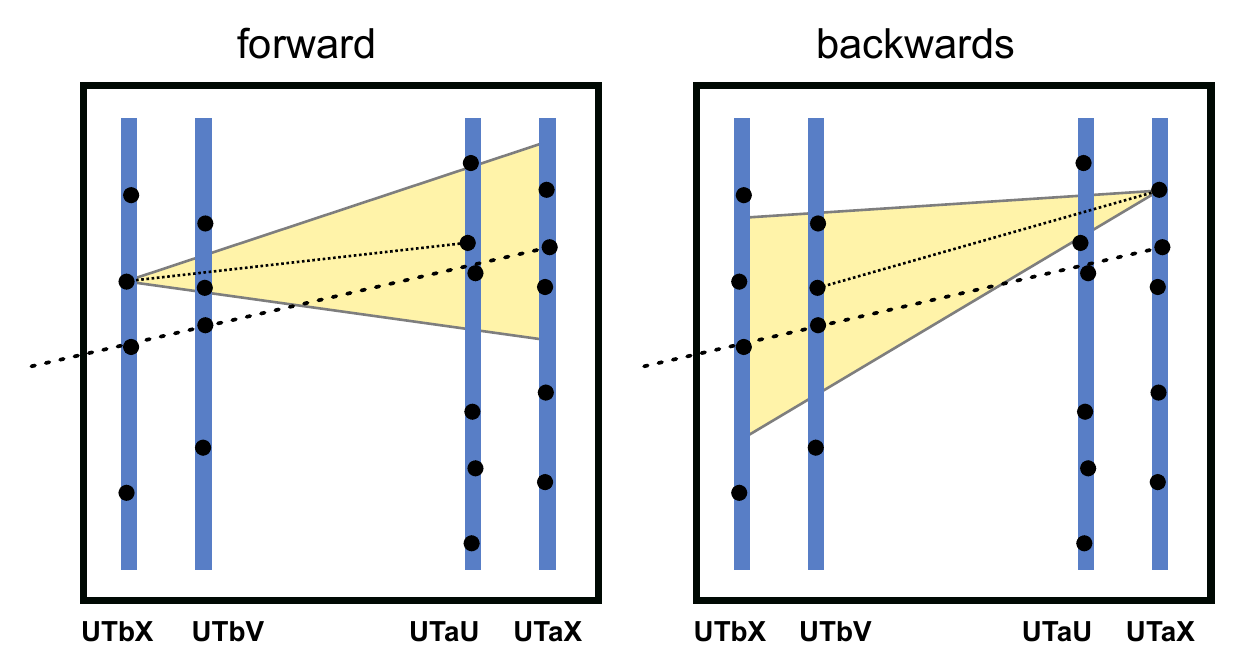}
  \caption{Tracklet finding kernel. Combinatorics between all 4 panels when searching for hits candidates to form a tracklet are shown. The fine dotted line represents the slope between the two first hits found in the first and third panels. The coarse dotted line represents the VELO track slope. A tolerance window defined by them is calculated to search for a tracklet.}
  \label{fig:find_best_hits}
\end{figure}

When a valid hit is found in the first panel, it is selected to be combined with a valid hit from the third panel. If a valid hit is also found in the latter the slope formed between them is calculated. The just calculated slope and the one of the VELO track are used to define a tolerance window in the second and fourth panels. Compatible hits are searched in these panels to form the final tracklet, as can be seen in Figure~\ref{fig:find_best_hits}. Finding a third hit is enough to from a tracklet, where a tracklet of four is preferred if it is found. The complexity of tracklet search is $O(n^3)$, as the search for third and fourth hits are not nested between them. The tracklet search is performed both in forward and backwards directions, where the same algorithm is applied changing the order of the panels. Forward and backwards search is merged into one single loop, where hits are searched first in forward direction and if no hits are found, the backwards direction is tested to find a tracklet.

The algorithm may be configured to use more than one window range, in this paper for one, three or five window ranges. Instead of looping independently over the ranges to find a tracklet, these are combined into one single loop, as if these were one single range. A pointer to a selected hit within a window range is used to iterate. The ranges are combined so the central one is used first, then its immediate neighbours. If five sectors were selected, the sectors in the extremes are searched the last. Forward and backward searches are combined, as we found this way of iterating over the hits to be faster than performing two separate searches for forward and backward direction, as thread divergence is removed. We parallelize the searches for every VELO track, where all the threads in a \emph{warp} will have to wait if a divergent branch is encountered in one of the threads. When we split the hit search into two loops, a divergent branch is introduced if different tracks are searching in forward and backward direction within a \emph{warp}. A small divergent branch is introduced at the beginning of the loop when combining the window ranges. This is done to set the pointer to the correct hit, which allows the \emph{warp} to run all tracks in a parallel fashion even if they diverge in both ranges or direction.

\emph{Compass} implements a configurable number of search hit candidates that will be considered. When a valid tracklet is found, if more than one candidate was configured, the next valid hits within the window ranges are tested to form a different tracklet. For every tracklet the $\chi^2$ fit of the track is obtained in combination with the VELO track. If more than one tracklet is found, we perform a selection favouring tracklets with 4 hits instead of 3, and with the lowest $\chi^2$ fit value. The algorithm keeps searching for a better tracklet according to the configured hit candidates value.

\emph{Compass} is parallelised over the VELO tracks, where each thread processes the tracklet search for each track. When processing the VELO tracks a similar filtering mechanism is applied as when searching for the window ranges explained in subsection~\ref{ssec:search_ut_windows}. It differs in the conditions to save a valid track, looking for the track to be withih UT acceptance, not backwards and to have at least one valid window range. Only the size of the window range is checked to be different from $-1$ to indicate a window range with at least one valid hit.

We also take advantage of the GPU shared memory to cache the window ranges, as these are accessed during the tracklet search. A shared memory array of size $num\_threads \times num\_panels \times size\_window\_range$ is used to accommodate all the window ranges in a block. As in the search window ranges kernel, we store the window ranges using a signed 16-bit type to save in memory. When processing a valid track, the window ranges for that track are copied to its correct position relative to the block size into shared memory, where only the pointers to the shared memory array are used afterwards. We found this to be faster in all the configurations and GPUs we tested.

When a final tracklet is selected as the best one, the found hits are stored and associated to its VELO track as a VELO+UT track. Alongside the hits, the charge of the particle, calculated from the momentum of the track from the $\chi^2$ fit, and the index of the track within the event are stored, obtained by atomic addition of the track number for this event.

\subsection{CPU implementation}\label{ssec:cpu_implementation}

We implement a CPU version of \emph{Compass} tracking to compare its computing performance against our baseline GPU implementation. To port the algorithm part of the structure of the algorithm is modified. The GPU specific optimizations are removed, which cannot be exploited in a non-GPU architecture. On the baseline GPU version we minimize thread divergence and store various structures into shared memory, whereas the impact of branches is minimized by design in a CPU architecture compared to a GPU architecture~\cite{lin2018gpu}~\cite{djenouri2017reducing}. We consider the impact of using shared memory and caching the window ranges in the ported version to be better managed by the large caches found in a modern CPU, compared to the ones in the GPUs. The computation of searching window ranges and tracklet finding is not split into separated kernels, where the window searches are calculated for every VELO track in-place before doing the tracklet search. We do so to benefit from cache locality, as the just calculated window ranges will be used by the tracklet search algorithm.

%% file: sections/06evaluation.tex
\section{Experimental evaluation}\label{sec:evaluation}

This section covers the performance and physics efficiency evaluation of our proposed algorithms. We have conducted multiple micro-benchmarks using different configurations for both the number of sectors and the number of candidates. 

\subsection{Experimental setup}\label{ssec:experimental_setup}

Four GPUs and a \mbox{x86-64} CPU were used for the benchmarks. Two consumer-grade GPUs of different generations and two server-grade GPUs are employed. A dual socket server-grade CPU is used for the \emph{Compass} tracking CPU implementation. The specifics of the hardware are detailed in Table~\ref{table:hardware}.

The software relies on CUDA 10.0 and \texttt{gcc} 7.3.0 under the \texttt{-O3} optimisation flag. The following compilation flags were used: \texttt{--use\_fast\_math} \texttt{--expt-relaxed-constexpr} and \texttt{--maxrregcount=63}. The use of those flags were beneficial for the overall execution time of our algorithm~\cite{nvidia_cuda}.

All the benchmarks use the same sets of Monte Carlo simulated events, generated using the LHCb simulation framework. Two different testbeds of events are evaluated: the \emph{minbias} set for throughput performance and the \emph{BsPhiPhi} to check reconstruction efficiency. The \emph{minbias} (minimum bias) set is a realistic simulation of the current expected physics, where data rate and therefore computing performance obtained with it match the realistically expected one. The \emph{BsPhiPhi} set contains more tracks from the rare decay $B_s \rightarrow \phi \phi$. This allows to determine the track reconstruction efficiency for these physically interesting decays with higher statistical significance. It is important to highlight that the same reconstruction efficiency can be achieved in both testbeds. However, we would need more \emph{minbias} samples to obtain the same number of tracks from the rare $B_s \rightarrow \phi \phi$ decay. Each set contains 1,000 events. For the throughput measurements, we iterate 40 times over the \emph{minbias} events to get a sustained throughput. Both server grade GPUs are set to ECC (Error-Correcting Code) memory disabled. The evaluation metrics shown in this paper correspond with the average value of 10 consecutive executions.

\begin{table}[hbt!]
\caption{GPU and CPU hardware employed for the evaluation. Two high-end consumer graphics cards (GeForce~GTX~1080Ti and GeForce~RTX~2080Ti), two server-grade cards (Tesla~T4 and Tesla~V100), and an Intel Xeon CPU are compared. We show the number of cores of each processor, where for the GPUs we count the CUDA cores only (no RT cores or Tensor cores are used in the benchmarks). We take the MSRP (manufacturer suggested retail price) for each hardware unit used here. The price for a single Intel Xeon CPU is shown, whereas for the benchmarks a dual socket server with two Intel Xeon CPUs is used. This is reflected in the price performance figure.}
\label{table:hardware}
\resizebox{0.98\linewidth}{!}{
\begin{tabular}{l||r|r|r|r|r|r}
\multirow{2}{*}{Unit}           & \multirow{2}{*}{\# cores} & Max freq.     & Cache       & DRAM    & TDP   & MSRP \\
               &                           & (GHz)         & (MiB - L2)  & (GiB)   & (W)   & (\$) \\
\hline
\hline
GeForce        & \multirow{2}{*}{3,584}   & \multirow{2}{*}{1.67}    & \multirow{2}{*}{2.75}    & 10.92       & \multirow{2}{*}{250}   & \multirow{2}{*}{699}       \\
GTX 1080 Ti    &                          &                          &                          & GDDR5       &                     &                         \\
\hline
GeForce        & \multirow{2}{*}{4,352}   & \multirow{2}{*}{1.54}   & \multirow{2}{*}{6}       & 10.92       & \multirow{2}{*}{250}   & \multirow{2}{*}{1,199}      \\
RTX 2080 Ti    &                          &                          &                          & GDDR5       &                     &                          \\
\hline
Tesla          & \multirow{2}{*}{2,560}   & \multirow{2}{*}{1.59}    & \multirow{2}{*}{6}       & 16          & \multirow{2}{*}{70}    & \multirow{2}{*}{2,350}      \\
T4             &                          &                          &                          & GDDR6       &                     &                          \\
\hline
Tesla V100     & \multirow{2}{*}{5,120}   & \multirow{2}{*}{1.37}    & \multirow{2}{*}{6}       & 16          & \multirow{2}{*}{250}   & \multirow{2}{*}{8,899}      \\
V100           &                          &                          &                          & HBM2        &                     &                          \\
\hline
Intel Xeon     & \multirow{2}{*}{20}      & \multirow{2}{*}{3.50}    & \multirow{2}{*}{25 (L3)} & 64          & \multirow{2}{*}{160}   & \multirow{2}{*}{2,145}      \\
E5-2678W v3    &                          &                          &                          & DDR4       &                     &           
\end{tabular}
}
\end{table}

\subsection{Compass tracking physics performance and throughput}\label{ssec:Compass_performance}

The computing performance of the algorithm is measured in terms of throughput of events per second. Different configurations of the algorithm are evaluated, taking measurements when looking into 1, 3, and 5 sectors and different number of hit candidates for 1 to 16 when looking for a better tracklet.

\begin{table*}[hbt!]
\centering
\caption{Comparison between searching in 1, 3 or 5 sector groups, and using 1 to 16 hit candidates. Two type of tracks are compared: long tracks and VELO+UT tracks. For each type of track, the track reconstruction efficiency and track clone rate achieved are presented. The obtained fake rate for each case is also shown.}
\begin{tabular}{c|c||cc|cc|c}
Number of                  & Number of  & \multicolumn{2}{c|}{Long tracks} & \multicolumn{2}{c|}{VELO+UT tracks} & \multirow{2}{*}{Fake rate} \\
sectors                    & candidates & reco. efficiency  & clone rate      & reco. efficiency & clone rate    &           \\
                           \hline
                           \hline
\multirow{5}{*}{1 sector}  & 1          & 71.91\%           & 0.36\%          & 61.88\%           & 0.32\%       & 7.73\%    \\
                           & 2          & 76.53\%           & 0.36\%          & 69.99\%           & 0.32\%       & 7.78\%    \\
                           & 4          & 79.09\%           & 0.31\%          & 74.31\%           & 0.32\%       & 7.70\%    \\
                           & 8          & 80.36\%           & 0.34\%          & 76.58\%           & 0.35\%       & 7.61\%    \\
                           & 16         & 80.52\%           & 0.34\%          & 77.04\%           & 0.35\%       & 7.52\%    \\
                           \hline
\multirow{5}{*}{3 sectors} & 1          & 84.70\%           & 0.39\%          & 66.87\%           & 0.32\%       & 7.64\%    \\
                           & 2          & 90.07\%           & 0.38\%          & 75.61\%           & 0.33\%       & 7.62\%    \\
                           & 4          & 93.31\%           & 0.35\%          & 80.32\%           & 0.32\%       & 7.52\%    \\
                           & 8          & 94.72\%           & 0.36\%          & 82.66\%           & 0.35\%       & 7.43\%    \\
                           & 16         & 94.94\%           & 0.36\%          & 83.19\%           & 0.35\%       & 7.33\%    \\
                           \hline
\multirow{5}{*}{5 sectors} & 1          & 85.23\%           & 0.39\%          & 67.10\%           & 0.31\%       & 7.70\%    \\
                           & 2          & 90.65\%           & 0.38\%          & 75.84\%           & 0.32\%       & 7.67\%    \\
                           & 4          & 93.89\%           & 0.35\%          & 80.52\%           & 0.32\%       & 7.56\%    \\
                           & 8          & 95.27\%           & 0.36\%          & 82.87\%           & 0.35\%       & 7.47\%    \\
                           & 16         & 95.49\%           & 0.36\%          & 83.40\%           & 0.35\%       & 7.38\%    \\
\end{tabular}
\label{table:efficiencies}
\end{table*}


The obtained physics efficiency is shown in Table \ref{table:efficiencies} for the long and VELO+UT tracks. We focus on the long tracks, as these are the preferred ones for analysis. Long tracks carry more information about the momentum resolution. We also analyse the VELO+UT tracks, as these are constructed with the two main inputs of the \emph{Compass} algorithm, VELO tracks and UT hits~\cite{lhcb2015measurement}. Note how for the 3 sector cases, when searching for more hit candidates, the physics efficiency improves. The biggest improvements are achieved in track reconstruction efficiency, where the clone rate increases by less than 0.1\% in all cases. Note how the reconstruction efficiency gains flattens when using more hit candidates. While the number of hit candidates is increased exponentially, the track reconstruction efficiency gains do not follow the same increase pattern, but the opposite. This behaviour matches our expectations, as in most of the cases, the best tracklet is found in the first set of hit candidates, and therefore, the subsequent ones do not yield a better hit tracklet as often. Calculating the subsequent tracklets has an impact on the throughput performance even if no better tracklet is found, where the physics performance does not improve. The fake rate decreases when using  more sectors and candidates, with differences in the range of 1\% across the whole scope of benchmarks. Note how the impact of both changing the sectors and candidates has little effect on the clone and fake rates, whereas it has a big impact in the reconstruction efficiency rate.

The reconstruction efficiency achieved when searching in one sector does not reach 90\% for long tracks nor 80\% for VELO+UT tracks for any number of hit candidates. These reconstruction efficiency does not meet the requirements for the LHCb UT reconstruction, and therefore, we discard the one sector configuration in the following analysis.

\begin{figure*}[hbt!]
  \centering
    \includegraphics[width=1\textwidth]{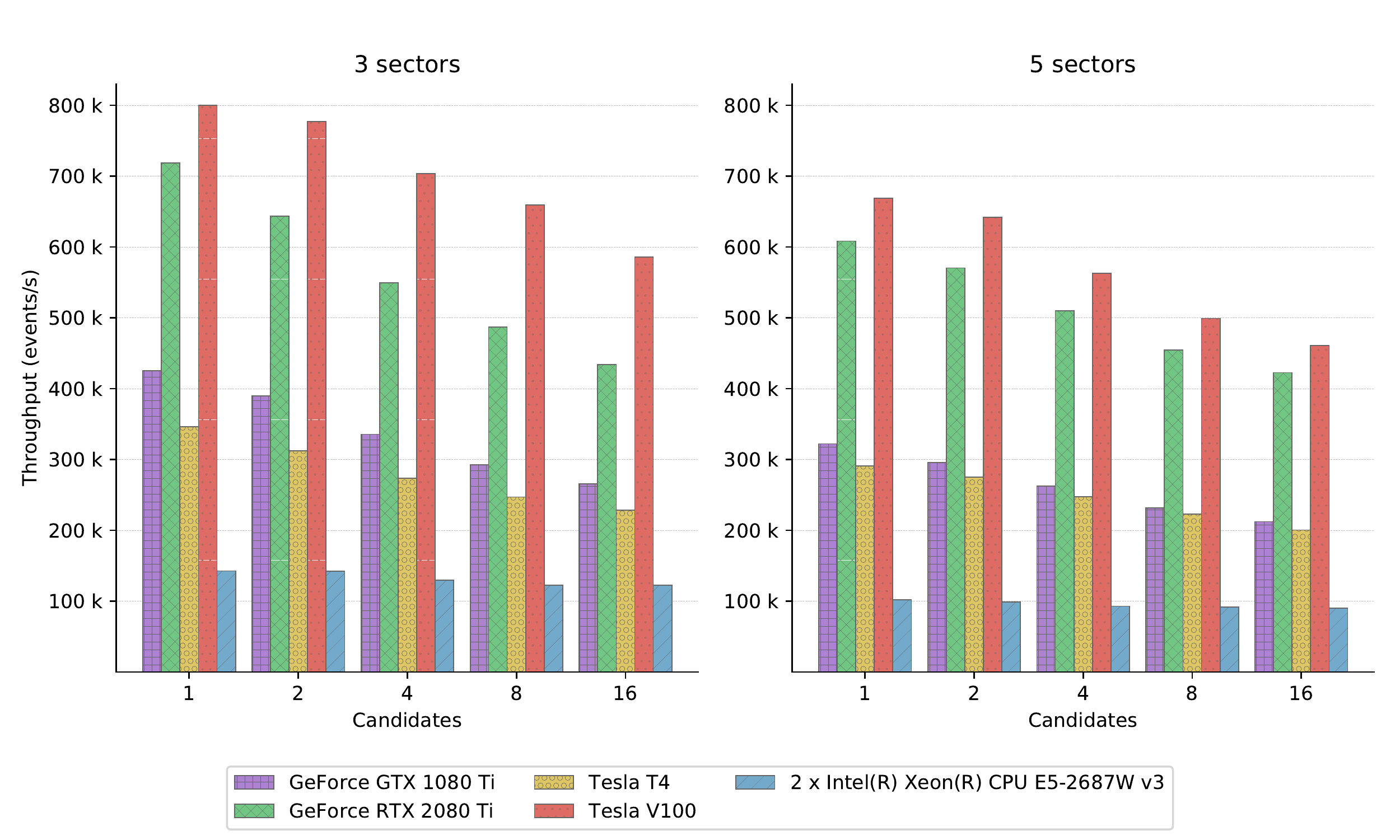}
  \caption{3 vs 5 sectors \emph{Compass} tracking comparison. Throughput comparison between the two consumer grade GPUs, two server grade GPUs and a dual socket Intel Xeon CPU, comparing with 1 to 16 number of hit candidates. The throughput shown here corresponds to running the \textit{Compass} algorithm. In the figure in the left we plot the throughput when looking for hits in 3 sectors. In the right figure, we depict the throughput when looking for hits in 5 sectors, adding an extra neighbour sector on each side with respect to the 3 sectors case.}
  \label{fig:throughput_sectors}
\end{figure*}

In Figure \ref{fig:throughput_sectors}, we plot the differences in throughput between all the configurations, using 3 and 5 sectors, and from 1 to 16 candidates. Note how searching for more candidates decreases the throughput, as it needs to iterate over more hits in a $O(n^3)$ algorithm to find a better hit tracklet. The performance degrades more when using more candidates, contrary to what we observed with the physics performance, where the gains were very small by doubling the number of candidates when using the bigger number of candidates. When searching for more hit candidates, the hit tracklet needs to be constructed, and their $\chi^2$ calculated, even if for most of the cases the last calculated hit tracklet does not improve over the previous one.

We highlight the difference in performance between the four evaluated GPUs devices. The 1080Ti and Tesla T4 have a comparable performance despite of the difference in terms of number of cores. We attribute the comparable performance between the two cards to the bigger cache size encountered in the Tesla T4 and its faster GDDR6 memory. The difference in thermal design power (TDP) is very significant, where the 1080Ti consumes $3\times$ more compared to the Tesla T4 to deliver a comparable throughput. The difference in performance between the 1080Ti / T4 compared to the 2080Ti is bigger than the difference found between the 2080Ti and the Tesla V100, with closer comparable performance when using 5 sectors compared to 3. Tesla V100 outperforms the rest of the GPUs due to its High Bandwith Memory (HBM) and increased number of cores, having double the number of cores compared to the T4, 15\% more compared to the 2080Ti, and 30\% more compared to the 1080Ti as show in Table~\ref{table:hardware}. One generation difference for the high-end consumer cards yields double the throughput for the 1080Ti compared to the 2080Ti for our algorithm.

Note the difference in performance for comparable physics efficiency on different results. We observe a comparable physics efficiency in the long tracks between the 5 sectors - 8 candidates case, and the 3 sectors - 16 candidates case. Taking the Tesla V100 as reference example, a difference in performance of roughly 15\% (500k vs 585k) is observed, whereas the difference in physics efficiency is below 1\%. The throughput differences change between the tested hardware for different number of candidates and sectors. Not that for comparable physics performance, the 5 sectors version performs better in throughput.


We port our \emph{Compass} tracking algorithm so that it runs on architectures other than the GPUs, to perform a cross-architecture tracking performance comparison. The CPU version differentiate from the GPU version in the implemented optimizations but computes the same algorithm and uses the same data layout and access patterns, as explained in~\ref{ssec:cpu_implementation}. OpenMP is used to parallelise over the events and tracks, following the same parallelisation scheme as in the GPU version. We ensure all cores are used in both the CPU and GPU versions for the comparison. Note how the parallelisation differs in the SIMD approach of the GPUs compared to the multi-threaded version of the CPUs, where the CPU version relies on the improvements made by the compiler due to the SoA data layout to exploit the SIMD capabilities of the CPU. The performance difference between a dual socket Intel Xeon CPU and the 1080Ti GPU and Telsa T4 is found to be up to 3$\times$ faster for the GPUs, up to 6$\times$ faster for the 2080Ti, and more than 6$\times$ faster for the V100. Note how the CPU version of the algorithm degrades less its performance compared to the GPUs when increasing both the number of sectors and candidates. We attribute this to the better branch prediction in the CPU, and the impact of divergent threads on the GPU, where the GPU runtime performance is affected more by the increased number of branches, and the work imbalance keeps warps active with low occupation, due to the increased number of candidates and sectors. 

\begin{figure}[hbt!]
  \centering
    \includegraphics[width=1\linewidth]{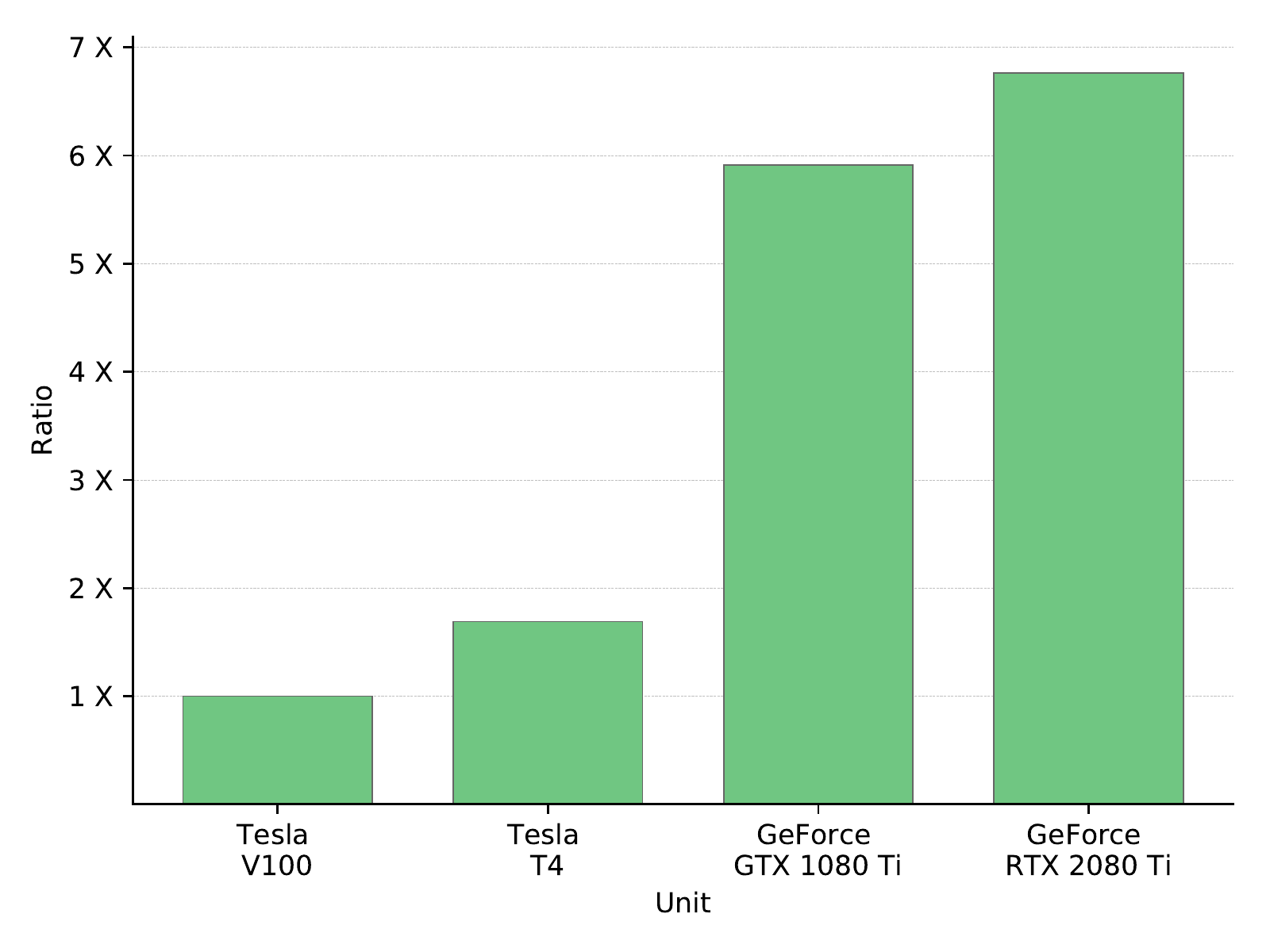}
  \caption{Price performance ratio for \emph{Compass} in GPU. All prices are factored to MSRP price indicated in Table~\ref{table:hardware}. We compare the price performance of the 5 sectors case, for the best physics efficiency case with 16 candidates.}
  \label{fig:price_performance_5_norm}
\end{figure}

In Figure \ref{fig:price_performance_5_norm}, we plot the price performance ratio for the different target GPUs. This figure shows the case for best physics performance with 5 sectors, using 16 hit candidates. It is normalised to the Tesla V100 and compares the other analysed hardware accelerators in terms of achieved speedup in terms of price/performance. Note how the price performance achieved for all the evaluated hardware is given for its MSRP\footnote{The prices shown in this paper are collected from those recommend by NVIDIA and Intel web site or Amazon.com otherwise.} with the prices shown in Table~\ref{table:hardware}.  Tesla V100 performs the worse in all the tested GPUs for its price performance, while it achieves the best throughput. Note the comparable price performance between the server grade Tesla GPUs compared to the consumer GPUs, where the consumer GPUs perform around 5$\times$ better than the server grade ones despite their differences in throughput. We note a 1.7$\times$ speedup between the Tesla V100 and the Tesla T4, and a 1.15$\times$ speedup between the 1080Ti and the 2080Ti, being the consumer grade GPUs close in price performance despite the 2080Ti doubling the 1080Ti in throughput. The achieved price performance speedup between the Tesla V100 and the 1080Ti is 5.9$\times$, and 6.7$\times$ for the 2080Ti. The 2080Ti obtains the best price performance due to the achieved high throughput and low unit price. The 2080Ti delivers a throughput close to the Tesla V100 with significant less price due the lack of some server-grade characteristics such as HBM or ECC memory. 


\subsection{UT decoding and tracking performance}\label{ssec:decoding_tracking_perf}

\begin{figure}[hbt!]
  \centering
    \includegraphics[width=0.95\linewidth]{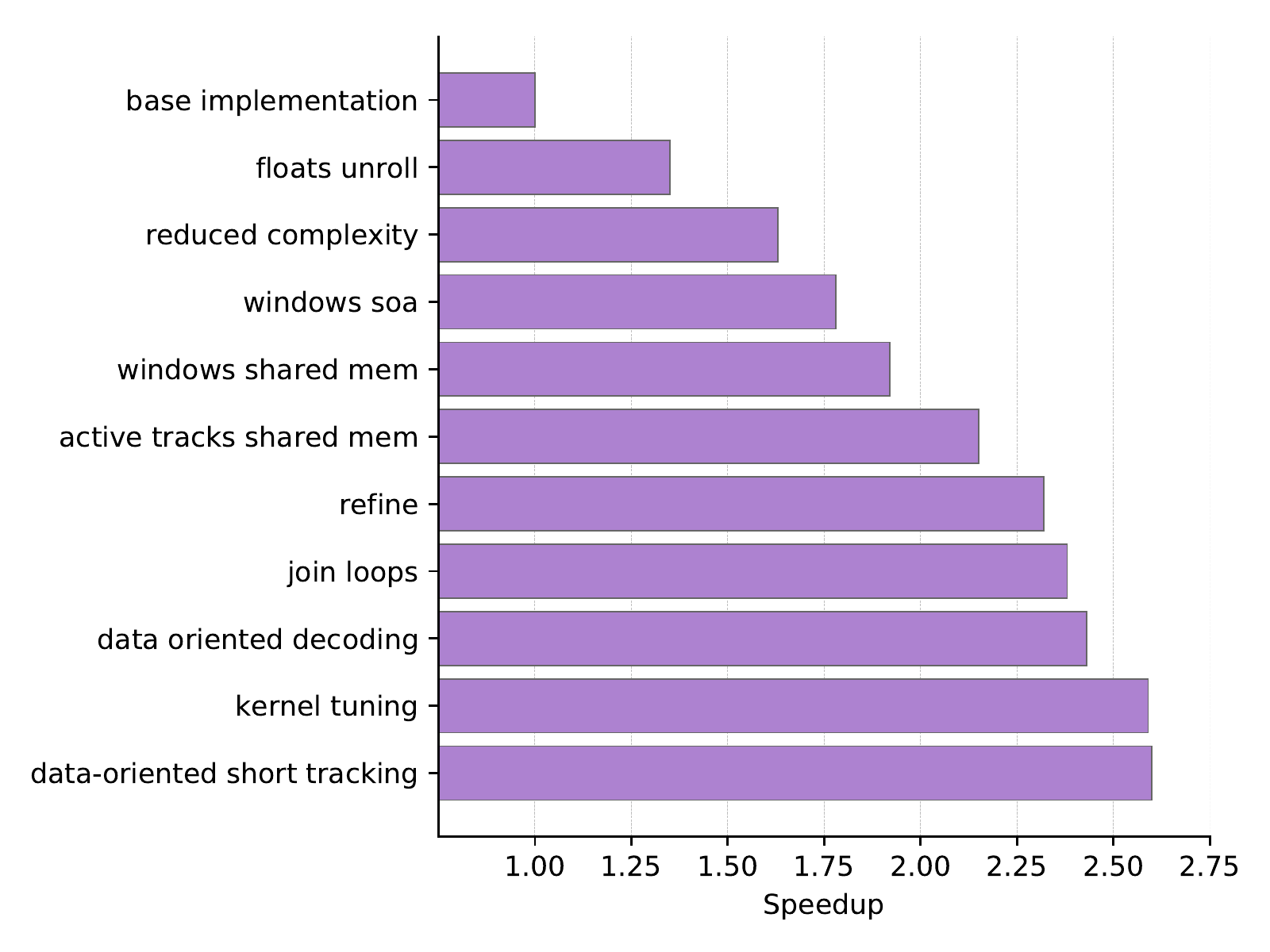}
  \caption{Incremental optimizations speedup. Speedup achieved after applying different optimizations to the baseline code. A maximum speedup of 2.6$\times$ is achieved in the final version, compared to the baseline implementation. Various small optimizations and changes are grouped into steps.}
  \label{fig:incremental_opt}
\end{figure}

In Figure \ref{fig:incremental_opt}, we show the speedup achieved for various iterations of optimizations, compared to the initial GPU implementation. Various small improvements and optimizations are grouped into the 11 steps presented in Figure~\ref{fig:incremental_opt}. We refer to the first working version that implements the main ideas of the algorithm as \emph{baseline implementation} and apply various optimization on top of it to achieve the final 2.6$\times$ speedup. For \emph{floats unroll}, we get the biggest improvement of 35\%. We first applied various small modifications to the algorithm, mainly changing all the floating point variables to single precision ones, unrolling some loops manually, and by giving compiler hints with the use of \texttt{\#pragma}. We note how the change from double to single precision does not affect the physics efficiency. We \emph{reduced the complexity} of window range search by splitting the algorithm in various kernels and re-writing the tracklet finding to be simpler to process when searching in more than one sector, to get a 28\% improvement. We improved the window ranges storage to be \emph{windows SoA} to get an extra 15\%, and configured it to store only one hit and the size of the window, sorting them to be efficient for our access pattern. We copied the \emph{windows to shared memory} to cache them and improve the access pattern when searching the tracklet. The speedup achieved by filtering the tracks in the shared memory array is 23\%, shown in \emph{active tracks shared mem}. When calculating the window ranges, we \emph{refine} the window by checking the hits in both extremes, instead of calculating all the window range validity in the tracking algorithm. We further reduced the complexity of the tracklet finding by \emph{joining the loops} and reducing thread divergence, where we got to 2.37$\times$. We grouped various small optimization to the raw bank decoding, making the data types smaller, aligned and more efficient to be a \emph{data oriented decoding}. We improved an extra 16\% by \emph{tuning the kernel} parameters of all the kernels in the decoding and \emph{Compass}, changing to multi-dimension kernels and changing how the kernels are parallelised. Finally, we reduced the memory footprint and made the copies faster by reducing further the data types, by storing types in signed 16-bit instead of 32-bits structures to get the final overall speedup of 2.6$\times$.

\begin{figure}[hbt!]
  \centering
    \includegraphics[width=0.95\linewidth]{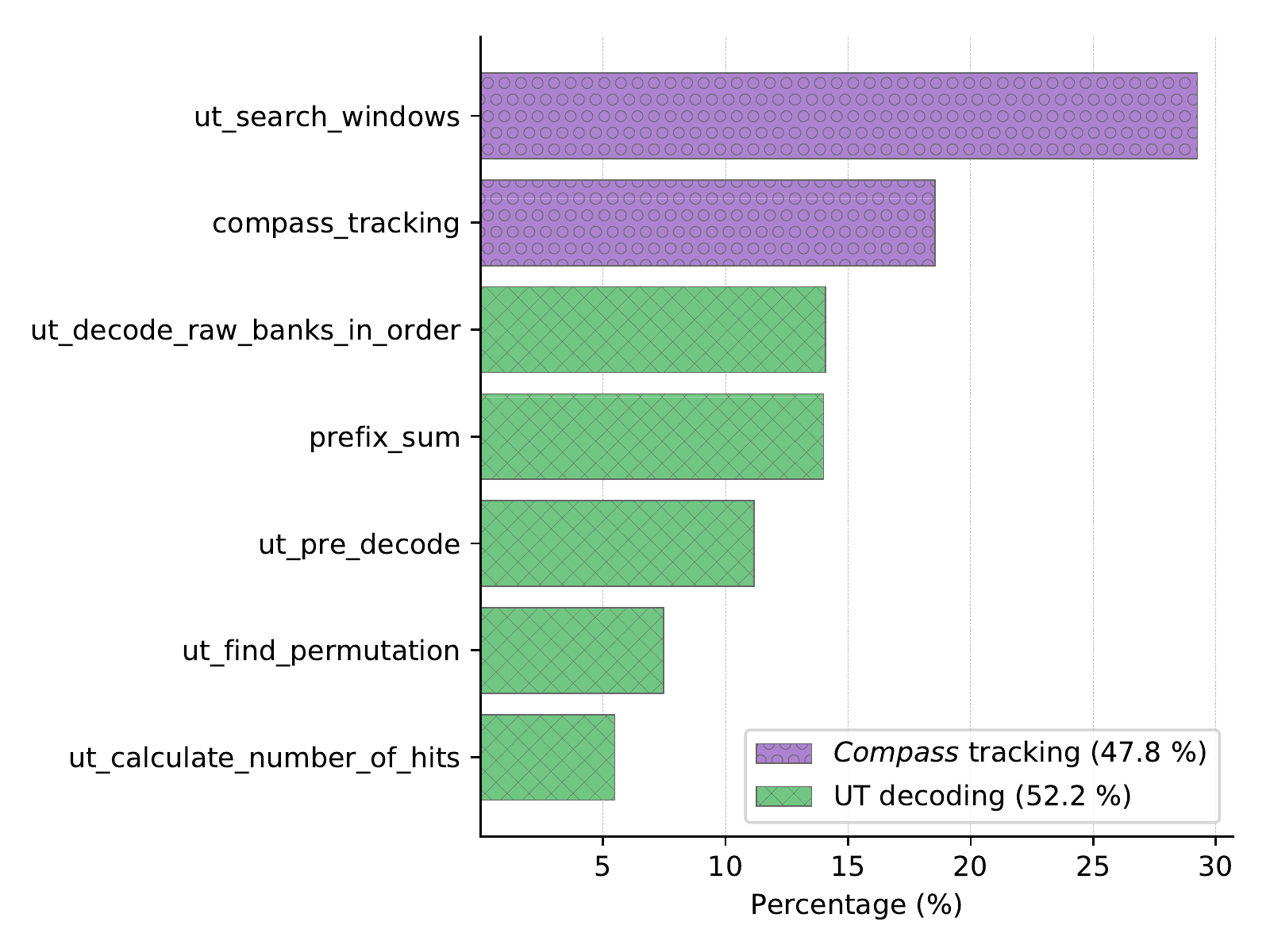}
  \caption{Kernels time contribution. Runtime distribution of all the kernels used to compute the decoding and \emph{Compass} algorithm. The best physics efficiency case is used here, with 5 sectors and 16 candidates for the NVIDIA 2080Ti case.}
  \label{fig:kernels_time}
\end{figure}

Figure \ref{fig:kernels_time} depicts the runtime distribution of both kernels used to perform the decoding and the kernels of the \emph{Compass} tracking algorithm. We show the distribution for the best physics case, 5 sectors - 16 candidates, where we encountered similar runtime distributions when using different configurations and different GPUs. Note how \emph{Compass} tracking runtime is dominated by the window searching algorithm compared to the tracklet finding. The refining of window ranges was moved from the tracklet finding to the window range search, increasing the time contribution of the kernel while improving the overall throughput. Note how the complete decoding of the UT hits accounts for more than half the time needed to compute the whole UT sequence.

\begin{figure}[hbt!]
  \centering
    \includegraphics[width=0.95\linewidth]{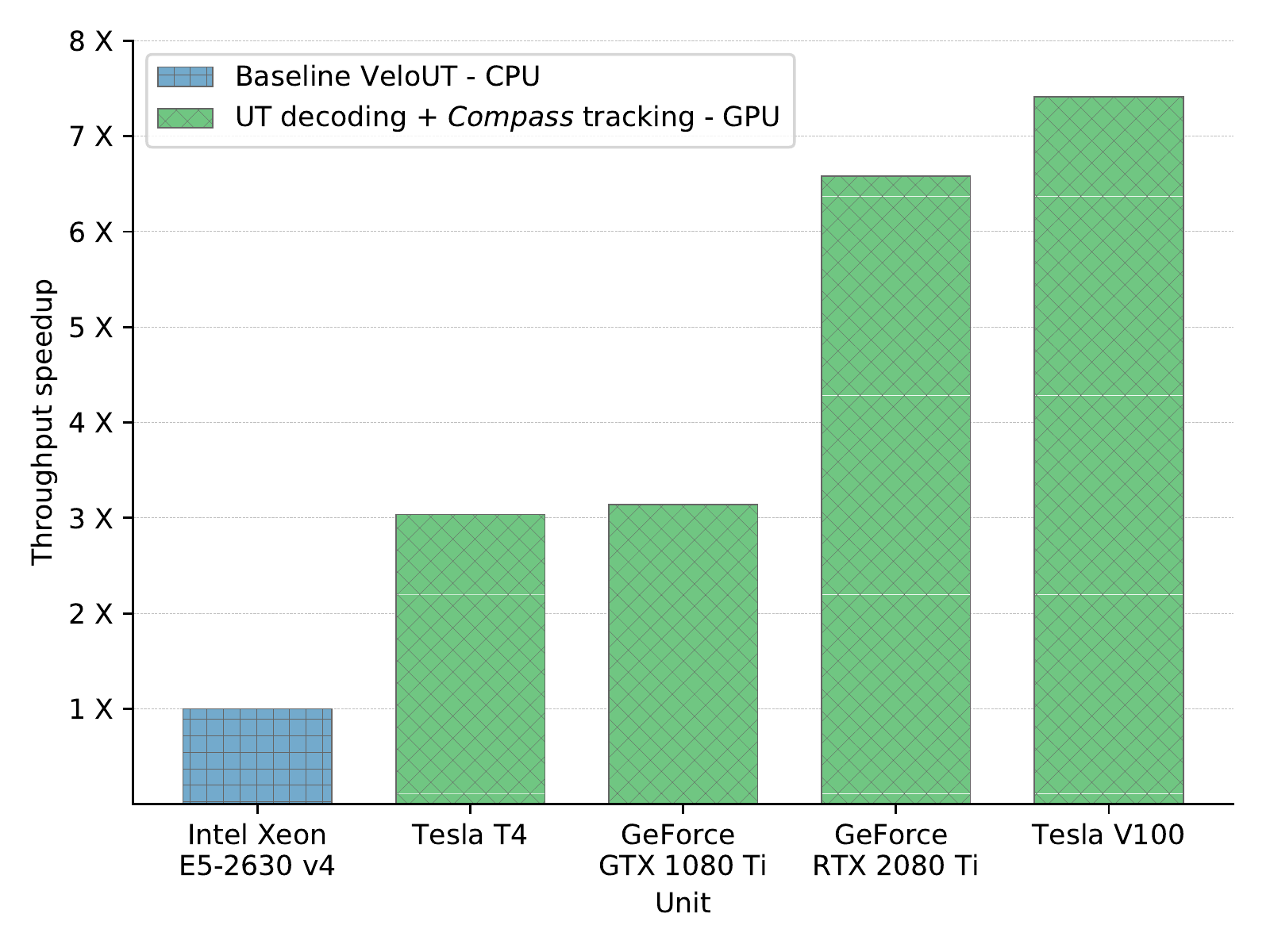}
  \caption{Basline LHCb vs GPU decoding + \emph{Compass} tracking throughput speedup comparison. Throughput speedup of the full UT chain of kernels, including the decoding and \emph{Compass} tracking, compared to the baseline LHCb CPU implementation as stated in Section~\ref{fig:cpu_baseline_speedup}. We compare the LHCb baseline (blue) with the \emph{Compass} over different GPUs (green).}
  \label{fig:cpu_baseline_speedup}
\end{figure}

Finally the complete implementation explained in this paper is shown, with the decoding and tracking in GPU compared to the equivalent algorithms found in LHCb baseline implementation. We acknowledge that the results compared here have changed and improved since the publication of these numbers in~\cite{de2018status} used for the comparison, where more recent results are not found or published. We set comparable conditions as those found in~\cite{de2018status}, where we apply the same Global Event Cut, which filters a selection of events, at the beginning of the chain, thus reducing the amount of processing the tracking algorithms need to do. We add data preparation kernels after the full UT chain is processed, in the form of a prefix sum and consolidation steps to leave the tracks in coalesced memory for the algorithms using UT tracks as input. The LHCb baseline implementation uses a Intel Xeon E5-2630 v4\footnote{This CPU differs from the one used for our benchmarks.}, which delivers a top throughput of 12,400 events per second for the full sequence.5 Combining the time contributions of the UT decoding and tracking for peak throughput yields the results shown in Figure~\ref{fig:cpu_baseline_speedup}. We compare these results to the full UT decoding and \emph{Compass} tracking presented in this paper. The throughput speedup shown corresponds to our \emph{Compass} implementation using the configuration for 5 sectors and 8 candidates. Both the Tesla T4 and 1080Ti achieve roughly a 3$\times$ speedup, where the latter performs slightly better than the T4. The 2080Ti achieves a speedup of 6.5$\times$ and the Tesla V100 achieves the best speedup at 7.4$\times$. We acknowledge that the physics results obtained in both implementations are comparable, but yield different results due to the different algorithms used.

%% file: sections/07conclusions.tex
\section{Conclusions}\label{sec:conclusions}

We have presented a new algorithm, \emph{Compass}, designed for parallel GPU architectures with focus to perform efficiently on GPUs. We designed our algorithm so that it maximises throughput processing on GPUs by being data-oriented, minimizing branching, reducing the memory footprint of the algorithm and taking advantage of the architectural characteristics of GPUs. 

We presented a SIMD parallel UT raw data decodification algorithm, data-oriented and optimized for GPUs. We demonstrated a new hit organization that stores hits in SoA, in a parallel and coalesced manner, where we sorted groups of hits into regions for fast decoding. We benefit from the new hit organization to search efficiently for sector regions, defining window ranges that indicate where compatible hits are found. We stored the windows efficiently for parallel architectures.

We designed \emph{Compass} to be configurable in both number of sectors to search for, and number of hit clusters to test for a tracklet. We showed the physics efficiency results when searching in one sector, proving it to yield too low reconstruction efficiency rate to be considered for performance benchmarks. We compared the performance for searching in three and five sectors, and tested with different number of hit candidates. We validated our algorithms with Monte Carlo simulated data to verify the physics performance of the results, getting comparable physics performance.

We developed a CPU tracking implementation and analysed our algorithm in different parallel architectures, focusing on GPU architectures and comparing them against the parallel CPU implementation of the same algorithm. We showed the differences in performance across the analysed hardware. We conclude that a physics performance close to $95\%$ in track reconstruction is achieved with various configurations of the algorithm, where a configuration using 5 sectors and 8 hit candidates yields a throughput of $231k$ events per second in the 1080 Ti, $222k$ in the Tesla T4, $454k$ in the 2080 Ti, $499k$ in the Tesla V100 and $92k$ in the dual socket Intel Xeon CPU, for the \emph{Compass} tracking. The $5\%$ of tracks that were not reconstructed correctly do not satisfy the assumptions and selections made in this algorithm. These are not due to computational precision, as has been verified switching from single to double precision obtaining the same results.

We consider this configuration to be the best trade-off for this algorithm considering the achieved physics efficiency and the performance. We compare with the baseline LHCb results for the full UT decoding and tracking, where our GPU implementation delivers up to 7.4$\times$ more throughput with the Tesla V100, and 6.5$\times$ when comparing with 2080Ti.

We plan to evaluate the possibilities of implementing further optimisations to the algorithm by exploiting various hardware capabilities of NVIDIA GPUs, such as the usage of Tensor and Ray Tracing cores. For the CPU implementation, vectorisation opportunities could be explored to further optimize the CPU implementation of the algorithm.